\documentclass[showpacs,amsmath,amssymb,twocolumn,prl,superscriptaddress]{revtex4-1}

\usepackage[dvips]{graphicx} 
\usepackage{amsfonts}
\usepackage{amssymb}
\usepackage{amscd}
\usepackage{amsmath}    
\usepackage{enumerate}
\usepackage{epsfig}
\usepackage{subfigure}
\usepackage{xcolor}
\usepackage{array}
\usepackage{amsthm}

\newcommand{\ket}[1]{\mbox{$\left| #1 \right\rangle$}}

\begin{document}
\title{Experimental passive round-robin differential phase-shift quantum key distribution}

\author{Jian-Yu Guan}
\affiliation{Department of Modern Physics and National Laboratory for Physical Sciences at Microscale, Shanghai Branch, University of Science and Technology of China, Hefei, Anhui 230026, China}
\affiliation{CAS Center for Excellence and Synergetic Innovation Center in Quantum Information and Quantum Physics, Shanghai Branch,  University of Science and Technology of China, Hefei, Anhui 230026, China}

\author{Zhu Cao}
\affiliation{Center for Quantum Information, Institute for Interdisciplinary Information Sciences, Tsinghua University, Beijing, 100084, China}

\author{Yang Liu}
\affiliation{Department of Modern Physics and National Laboratory for Physical Sciences at Microscale, Shanghai Branch, University of Science and Technology of China, Hefei, Anhui 230026, China}
\affiliation{CAS Center for Excellence and Synergetic Innovation Center in Quantum Information and Quantum Physics, Shanghai Branch,  University of Science and Technology of China, Hefei, Anhui 230026, China}

\author{Guo-Liang Shen-Tu}
\affiliation{Department of Modern Physics and National Laboratory for Physical Sciences at Microscale, Shanghai Branch, University of Science and Technology of China, Hefei, Anhui 230026, China}
\affiliation{CAS Center for Excellence and Synergetic Innovation Center in Quantum Information and Quantum Physics, Shanghai Branch,  University of Science and Technology of China, Hefei, Anhui 230026, China}

\author{Jason S. Pelc}
\affiliation{Edward L. Ginzton Laboratory, Stanford University, Stanford, California 94305, USA}

\author{M. M. Fejer}
\affiliation{Edward L. Ginzton Laboratory, Stanford University, Stanford, California 94305, USA}

\author{Cheng-Zhi Peng}
\affiliation{Department of Modern Physics and National Laboratory for Physical Sciences at Microscale, Shanghai Branch, University of Science and Technology of China, Hefei, Anhui 230026, China}
\affiliation{CAS Center for Excellence and Synergetic Innovation Center in Quantum Information and Quantum Physics, Shanghai Branch,  University of Science and Technology of China, Hefei, Anhui 230026, China}

\author{Xiongfeng Ma}
\email{xma@tsinghua.edu.cn}
\affiliation{Center for Quantum Information, Institute for Interdisciplinary Information Sciences, Tsinghua University, Beijing, 100084, China}

\author{Qiang Zhang}
\email{qiangzh@ustc.edu.cn}
\affiliation{Department of Modern Physics and National Laboratory for Physical Sciences at Microscale, Shanghai Branch, University of Science and Technology of China, Hefei, Anhui 230026, China}
\affiliation{CAS Center for Excellence and Synergetic Innovation Center in Quantum Information and Quantum Physics, Shanghai Branch,  University of Science and Technology of China, Hefei, Anhui 230026, China}

\author{Jian-Wei Pan}
\affiliation{Department of Modern Physics and National Laboratory for Physical Sciences at Microscale, Shanghai Branch, University of Science and Technology of China, Hefei, Anhui 230026, China}
\affiliation{CAS Center for Excellence and Synergetic Innovation Center in Quantum Information and Quantum Physics, Shanghai Branch,  University of Science and Technology of China, Hefei, Anhui 230026, China}

\begin{abstract}
In quantum key distribution (QKD), the bit error rate is used to estimate the information leakage and hence determines the amount of privacy amplification --- making the final key private by shortening the key. In general, there exists a threshold of the error rate for each scheme, above which no secure key can be generated. This threshold puts a restriction on the environment noises. For example, a widely used QKD protocol --- BB84 --- cannot tolerate error rates beyond 25\%. A new protocol, round-robin differential phase shifted (RRDPS) QKD, essentially removes this restriction and can in principle tolerate more environment disturbance. Here, we propose and experimentally demonstrate a passive RRDPS QKD scheme. In particular, our 500 MHz passive RRDPS QKD system is able to generate a secure key over 50 km with a bit error rate as high as 29\%. This scheme should find its applications in noisy environment conditions.
\end{abstract}

\maketitle

The uncertainty principle  guarantees that whenever an eavesdropper, Eve wants to learn key information in the quantum channel,
she would inevitably introduce disturbances, which could be detected by the two authorized parties, Alice and Bob.
In reality, the quantum channel may suffer from environment disturbance, which could cause errors and even more vitally conceal Eve's attack.

The amount of leaked key information, which is quantified by a phase error $e_{p}$, can be inferred from the channel disturbance, which is quantified by a bit error $e_{b}$. The final key rate is given by \cite{SHORPRESKILL2000PRL},
\begin{equation}\label{Rebep}
R \ge 1-H(e_b)-H(e_p),
\end{equation}
where $H(e)=-e\log e -(1-e)\log(1-e)$ is the binary Shannon entropy function. The bit error can be directly computed from the experimental data, whereas the phase error needs to be estimated or bounded. In the BB84 protocol with strong symmetries, one can show that $e_{p}=e_{b}$ in the long key length limit. In other protocols, normally there is a relation between the two error rates. In the end, when the error rate $e_b$ goes beyond some threshold level, no secure key can be generated. For example, with the Shor-Preskill security proof \cite{Lo1999Science,SHORPRESKILL2000PRL}, the BB84 protocol can maximally tolerate 11\% error rate. For any security analysis, a simple intercept-and-resend attack \cite{Bennett:BB84:1984} shows that the BB84 protocol cannot tolerate more than 25\% error rate. This threshold puts a stringent requirement on the system environment, which makes some practical implementations challenging.

\begin{figure} [hbt]
\centering
\resizebox{8.5cm}{!}{\includegraphics{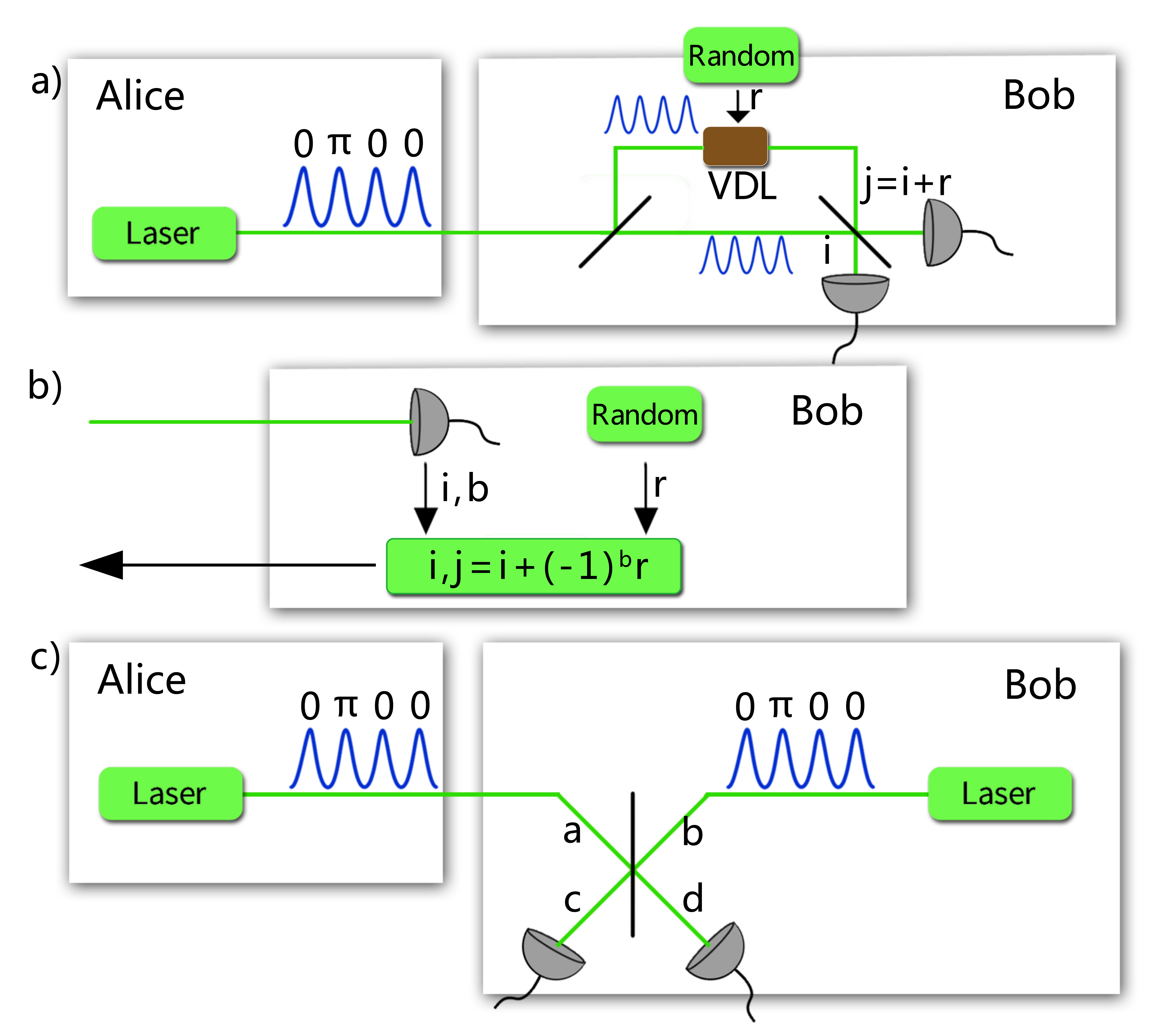}}\\
\caption{(a) Original RRDPS scheme \cite{sasaki2014practical}. VDL stands for variable delay line. Bob splits the received signals into two paths and applies a variable delay $r$ to one of the paths. 
(b) Equivalent model. Bob obtains a click at position $i$, generates two random numbers $r\in \{1,\cdots,L-1\}$ and $b\in \{0,1\}$ to obtain $j=i+(-1)^br$, and publicly announces $i$ and $j$ to Alice. (c) Passive RRDPS scheme. Bob uses a local laser to generate an $L$-pulse reference, which interferes with Alice's $L$-pulse signal. Bob then records the coincidence clicks.} \label{FigScheme}
\end{figure}

Recently, Sasaki et al.~proposed a round-robin differential phase-shift (RRDPS) QKD protocol \cite{sasaki2014practical}.
The sender Alice encodes a random phase, chosen from $\{0, \pi\}$, on each of $L$ pulses, with an average photon number of $\mu$. Upon receiving the $L$-pulse block, the receiver Bob implements a single-photon interference with an Mach-Zehnder interferometer (MZI), as shown in Fig.~\ref{FigScheme}a. The key point is that Bob can randomly adjust the length difference of the two arms of the MZI. After obtaining a detection click, Bob first identifies which two pulses interfere and then announces the corresponding indices $i$, $j$ to Alice. Alice can derive the relative phase between the two pulses as the raw key, and Bob can record the raw key from the measurement results. The phase error rate $e_{p}$ depends only on the number of photons in the $L$-pulse signal and $L$, not the bit error rate $e_{b}$. By setting a large enough $L$, the phase error rate tends to 0 and hence the scheme can tolerate up to 50\% bit error rate according to Eq.~\eqref{Rebep}.

In the protocol, Bob needs to randomly adjust the length difference between the two arms of the MZI, from $1$ to $L-1$ pulse periods. Based on the current technology, however, it is challenging to quickly change the length difference of the two arms in an MZI. The main adjust-delay method is to utilize optical switches, which cannot provide both high speed and low insertion loss simultaneously.

In this Letter, we propose an alternative scheme that passively chooses two pulses to interfere with the same bit error tolerability. When Bob receives a block from Alice, he prepares a local $L$-pulse reference in plain phases, i.e., all phases are encoded at phase 0. This $L$-pulse reference interferes with the $L$-pulse signal sent by Alice on a beam splitter, as shown in Fig.~\ref{FigScheme}c. For each block, Bob records the status of his two detectors with timestamps, $i$ and $j$.

If Bob's reference is in phase with Alice's signal, the whole setup is essentially a huge MZI. Any detection signal at time slot $i$ will tell the phase difference between $i$ and the phase reference. Then the encoding bit value can be revealed to Bob. Here, Bob requires a phase reference from Alice, which may require complicate frequency comb technology \cite{phase-lock:Shelton2001}.


The phase reference is not necessary requirement for our scheme though. If Bob's phase reference is random comparing to Alice's signal, the interference is no longer a Mach-Zehnder type but a Hong-Ou-Mandel type \cite{Hong-Ou-Mandel}. Let us consider a simple case when both Alice and Bob each has exactly one photon in their $L$-pulse trains. The states of Alice and Bob can be represented by
\begin{equation}
\begin{aligned}
\frac{1}{\sqrt{L}}\sum\limits_{i=1}^L (-1)^{s_i}a_i^{\dagger}\ket{0}, \quad \frac{1}{\sqrt{L}}\sum\limits_{i=1}^L b_i^{\dagger}\ket{0}
\end{aligned}
\end{equation}
respectively, where $s_i\in \{0,1\}$ designates the phase of Alice's $i$-th pulse. Since there are two photons in a block, one from Alice and one from Bob, Bob would obtain at most two detection clicks. He post-selects to choose the block where there are exactly two detections and announces their positions $i$ and $j$ (if $i=j$, the detection result is discarded). The raw key is the relative phase between these two pulses in the $L$-pulse signal. Alice can derive this phase difference from her record.

After the interference and Bob's post-selection, the quantum state at the two detectors becomes one of
\begin{equation}
\begin{aligned}
 & (1-(-1)^{s_i+s_j})d_i^{\dagger}c_j^{\dagger}\ket{0}, \quad  (1-(-1)^{s_i+s_j})c_i^{\dagger}d_j^{\dagger}\ket{0},  \\
& (1+(-1)^{s_i+s_j})c_i^{\dagger}c_j^{\dagger}\ket{0}, \quad (1+(-1)^{s_i+s_j})d_i^{\dagger}d_j^{\dagger}\ket{0},
\end{aligned}
\end{equation}
where $c_i^{\dagger}$ and $d_i^{\dagger}$ are the creation operators at the two detectors respectively, as shown in Fig.~\ref{FigScheme}c. This means if Alice's pulses $i$ and $j$ have the same phase, i.e., $s_i=s_j$, the two clicks should be triggered by the same detector. While if Alice's pulses $i$ and $j$ have different phases, the two clicks should be triggered by different detectors. Thus Bob can derive the relative phase by comparing the measurement results of the $i$-th and $j$-th pulses.

For the security analysis, we show that in the single photon case, our protocol is equivalent to an intermediate model shown in Fig.~\ref{FigScheme}b \cite{sasaki2014practical}, which is then equivalent to the RRDPS protocol. Thus the phase error $e_{p}$ is also bounded by $1/(L-1)$ as in RRDPS.
Bob post-selects the block where two clicks happen at $i$ and $j$, but he cannot distinguish whether the photon causing the click belongs to the signal (Alice) or to the reference (Bob).
Suppose Bob's photon is at $i$; the other case is similar. Since the $L$-pulse reference of Bob has a symmetry among all pulses, the $L-1$ possible positions of Bob's photon $i$ (excluding the position of Alice's photon, $j$) have the same weight. Bob has passively chosen a random shift $r=j-i$ between the clicks $i$ and $j$, which is equivalent to the active shift in the raw model, as shown in Fig.~\ref{FigScheme}a. We give a strict proof of this equivalence in the Supplemental Material, by showing that for any single photon input to Bob, the output, which is the distribution of the detection event $(i,j)$, remains the same for both our protocol and the raw model.


\begin{figure}
\centering
\resizebox{8.5cm}{!}{\includegraphics{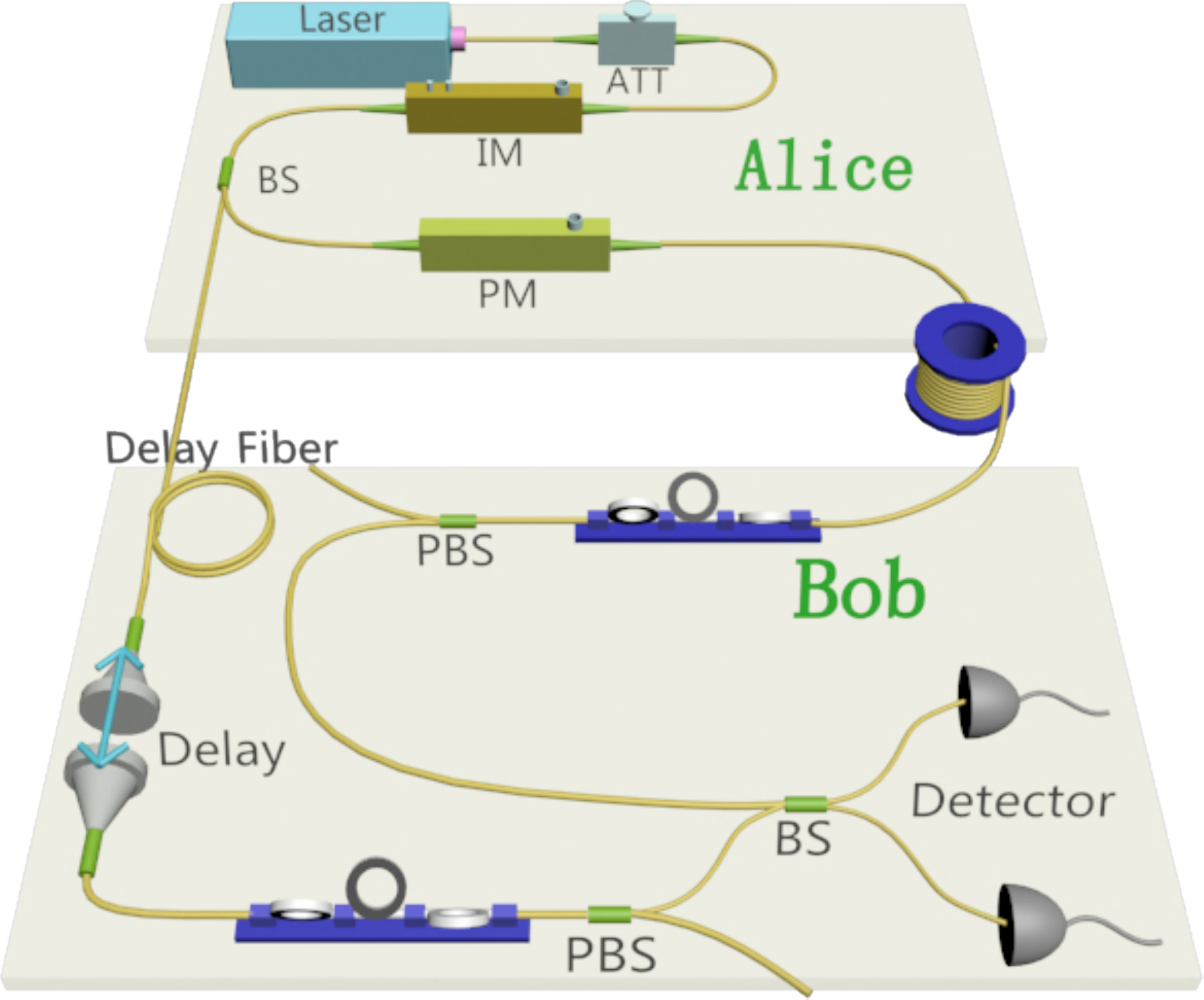}}\\
\caption{Experiment setup. ATT: attenuator; IM: intensity modulator; BS: polarization maintaining beam splitter; PM: phase modulator; Delay: optical adjustable delay line; PC: polarization controller; PBS: polarization beam splitter (single mode to polarization maintaining); Det: detector. The attenuation on Bob's side is realized by a polarization controller and a polarization beam splitter.} \label{FigExp}
\end{figure}

In practice, a single-photon state source is often replaced by a weak laser pulse, which can be described by a coherent state. Alice generates a coherent state pulse, randomizes its phase, and divides it into a series of weaker coherent state pulses using, say, beam splitters. Alice can also generate the pulse train directly, say by modulating a continuous-wave laser, with the same phase, which we called overall phase. This state preparation is the same for Bob. When the overall phase is randomized, it is shown that the state of the whole pulse train can be described by a statistical mixture of Fock states, whose photon number follows a Poisson distribution \cite{Lo:PRL:2005}. Similar to the single-photon case, Alice's key information is encoded into the relative phases between pulses.

In this coherent-state scenario, it is possible to have multi-photon components in both Alice and Bob's respective pulse trains, which will invariably alter Bob's post processing strategy. If Bob gets two or more detector clicks in a block, he randomly chooses two timestamps  $i$ and $j$ of detector clicks and announces them. Otherwise, he discards the result. In this way, Bob can figure out the phase difference between $i$ and $j$ as in the single photon protocol. By dividing into cases, one can bound $e_{p}$ for the coherent-state protocol. Detailed analysis can be found in Supplemental Material.

Meanwhile, the multi-photon components may cause a large inherent bit error rate. Imagine the case where Alice sends nothing (or photons are lost in the channel) and Bob sends 2 photons, it might result in a false conclusive detection event. The bit error rate in this case is clearly 50\%. Since the probability of two single photons from each side (which does not have any inherent error) is the same as the probability of multi-photon from one arm and nothing from the other, the total inherent bit error rate will be 25\%.

Our passive RRDPS scheme has three possible types of implementations: single-photon case, phase-locked weak coherent state, phase-randomized weak coherent sate. The third case is the most practical, which does not needs any fancy phase-locking technology, single photon source or high-speed optical switches. We then provide a proof-of-principle experiment demonstration for the third case.



The experiment setup is shown in Fig.~\ref{FigExp}. On Alice's side, a tunable extra cavity diode laser with a central wavelength of 1550.14 nm and a line-width of 50 kHz is modulated into a pulse train with a repetition frequency 500 MHz. A beam splitter (BS) is used to separate the pulse train into two beams, one is for Alice's encoding and the other one is sent to Bob as a reference. Alice encodes random \{0, $\pi$\} phase into each individual pulse of the pulse train with a 10 GHz modulator, driven by a pulse pattern generator. The pulse pattern generator's random signal is generated beforehand by a quantum random number generator. Before sending the pulse train to the channel, Alice attenuates the average photon number per pulse into 0.004. The signal light goes through the channel of a fiber spool to Bob.



On Bob's side, he first attenuates his reference pulse intensity into an average photon number $\mu=0.004$ per pulse, and then interferes with the reference pulse on a BS.
Before the BS, a tunable fiber delay line and some fixed fiber delay are used to guarantee that the two pulse trains arrive at the BS simultaneously. Meanwhile, two polarization controllers and polarization beam splitters are used to make the two beams' polarization identical.



The output ports of the BS are led to two up-conversion single photon detectors. The up-conversion
detector uses sum-frequency generation in a periodically poled lithium niobate (PPLN) with a 1.94 $\mu$m pump beam to convert the telecom-band photons to 860 nm, where they are detected by a silicon single-photon detector. This scheme benefits from the high detection efficiency and short dead time of the silicon detector, and the long-wave pump technology \cite{Pelc2011} as well as the volume bragg gratings (VBG) help to reduce the noise dramatically \cite{Shentu:upconversion:13}.
The detectors used in our experiment both have efficiencys larger than 14\%, a dead time less than 80 ns, and a dark count 500 Hz.

We utilize a time digital converter (TDC) to record the detection signal. The TDC has a timing resolution of 160 ps and is synchronized with the pulse pattern generator by sharing the same clock. The TDC will time tag and memorize all the events and sent to a PC for analysis.


The final key rate formula is similar to the BB84 protocol \cite{SHORPRESKILL2000PRL},
\begin{equation}\label{finalkeyrate}
K=N(1-H_{PA}-H_{EC}),
\end{equation}
where $N$ stands for the length of sifted key and $H_{EC}= f \cdot H(e_{b})$ accounts for the cost of error correction. Here, denote $e_{b}$ to be the bit error rate, and $f$ to be the error correction efficiency. For simplicity, we use $f=1$ in the following postprocessing.

The privacy amplification cost is $H_{PA}=H(e_{p})\times(1+1.98 \sqrt{s/N})$ where $e_p$ is the phase error rate and the second factor accounts for finite key effects. Here $s$ comes from the security parameter $2^{-s}$, a typical value of which is 100. To estimate the phase error rate, we set a proper photon number threshold $v_{th}$, which is a parameter to be optimized. For an $L$-pulse signal containing more than $v_{th}$ photons, we assume these photons as tagged and the corresponding $e_p=1/2$, that is, Eve can get all information about them. For a signal containing less than $v_{th}$ photons, we can effectively bound the leaked information by estimating its phase error rate (see Part II of Supplemental Material). Let $e_{src}(v_{th})=Pr(n>v_{th})$ be the probability of this, where $n$ is the photon number of the $L$-pulse signal.
The phase error is calculated by
\begin{equation}\label{Eq:phaseoriginal}
\begin{aligned}
e_{p}=\frac{e_{src}}{Q}+(1-\frac{e_{src}}{Q})\frac{1-(\frac{L-3}{L-1})^{v_{th}}}{4}+\frac{\frac{m}{M}}{2(1-\frac{m}{M})},\\
\end{aligned}
\end{equation}
where $Q$ is the gain of the experiment given by $N_{em}/N$.  Here $N_{em}$ is the total number of blocks, $N$ is the number of blocks after Bob's post-selection, $m$ is the total number of photon counts of one detector and $M$ is the total number of pulses.

The three additive terms in the phase error correspond to the probability of more than $v_{th}$ photons, the probability of less than $v_{th}$ photons and the probability that two or more photons simultaneously enter the same detector at the same timestamp.
Note that one factor 2 in the denominator of $(1-(\frac{L-3}{L-1})^{v_{th}})/4$ is because the phase error rate is 0 when the two clicks that Bob announces are both from the signal or both from the reference, and this probability is at least as large as the probability that one such click comes from the signal and the other from the reference. By choosing a proper value of $v_{th}$, one can minimize the cost of privacy amplification. The detailed discussion is referred to the Supplemental Material.



\begin{figure}[h]
  \centering
\resizebox{8.5cm}{!}{\includegraphics{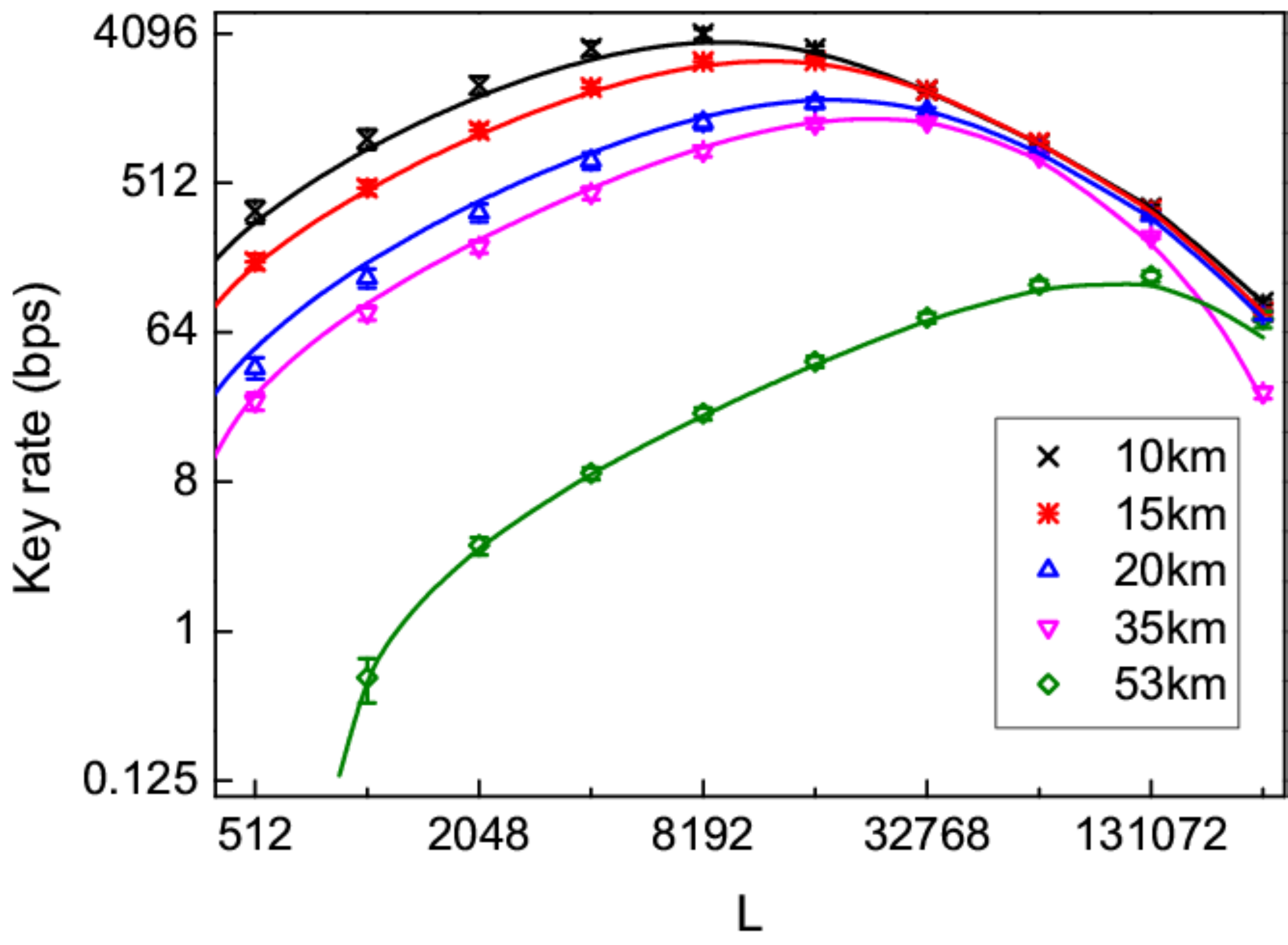}}\\
\caption{The dependence of the key rate on the block size $L$ at various distances. For each distance, we repeat the experiment by 10 times and take the standard deviations in 10 trials as the error bar.} \label{FigKeyAll}
\end{figure}

The final key rate depends on the block length $L$. Given the laser intensity of every pulse and the transmission distance, there exists an optimal $L$ for the key rate. On Alice's side, instead of setting a fixed $L$, we modulate the CW laser to form a continuous sequential pulse train, like the DPS QKD experiment \cite{Hiroki:QKD40dB:2007, DPS2009PRL}. During the postprocessing step, we can choose an optimal $L$ by maximizing the final key rate, as shown in Fig.~\ref{FigKeyAll}.

The experimental parameters with the optimal $L$ are listed in Table \ref{KeyRate}.

\begin{table}[h]
\centering
\caption{List of optimal $L$, $v_{th}$, $e_b$, and $e_{ph}$ for various distances.} \label{KeyRate}
\begin{tabular}{ccccccccccccc}
\hline
\hline
  d(km) & $\mu$ & optimal L & $v_{th}$ & $e_b$ & $e_{ph}$\\
  \hline
  10 & 0.004 & 8192 & 57 & 0.275 & 0.00359\\
  15 & 0.004 & 16384 & 99 & 0.271 & 0.00311\\
  20 & 0.004 & 16384 & 100 & 0.283 & 0.00312\\
  35 & 0.004 & 32768 & 179 & 0.276 & 0.00278\\
  53 & 0.004 & 131072 & 625 & 0.312 & 0.00240\\
\hline
\hline
\end{tabular}
\end{table}

The above analysis does not consider the dead time. We discount its effect by post-selecting: immediately after
one detector click, we effectively disable the other detector within one dead time period by post-selecting out this period.
The exact treatment is referred to the Supplemental Material.


In summary, we demonstrate a passive scheme to substitute the original RRDPS protocol and our system can distill a secure key with a bit error rate of 28\% in the lab. With our scheme, one can easily achieve a large number $L$ (say, $L=2^{14}$) of pulse train in experiment. In the original scheme, the pulse train length $L$ needs to optimized before the experiment, which requires a precise calibration of the system. In our scheme, on the other hand, the parameter $L$ can be decided during postprocessing step, which has an advantage in the case with large environment fluctuations.

The inherent error can be removed by a post-selecting technique \cite{MXF:MIQKD:2012} combined with the recently developed discrete-phase-randomization scheme for the coherent states \cite{Discrete-phase-randomized}. In principle, with certain modifications, such technique can be used in our scheme. This is an interesting prospective research project. Meanwhile, the inherent error can be removed by using phase-locked coherent state. Note that if Alice sends a strong laser pulse to Bob as reference and Bob directly uses it as for interference, Eve may implement a man-in-the-middle attack. One solution to remove this potential threat is to utilize frequency comb based frequency distribution technology \cite{Frequency-Metrology:920km}.

In future, a field test of the passive RRDPS scheme with two independent lasers
can be realized by the technology developed in a recent QKD experiment \cite{field-test:Yanlin}.
With low-jitter and high-efficiency single photon detectors \cite{superconducting-93}, a much higher secure
key rate with a 10 GHz clock rate system \cite{Hiroki:QKD40dB:2007} can be achieved.

\textbf{\subsection*{Acknowledgments}}
The authors would like to thank Yan-Lin Tang, Xiao Yuan and Zhen Zhang for enlightening discussions. This research has been supported by the National Basic Research Program (under Grant No. 2011CB921300, 2013CB336800 and 2011CBA00300), the National Natural Science Foundation of China, the Chinese Academy of Science, and the Shandong Institute of Quantum Science \& Technology Co., Ltd.



\bibliographystyle{apsrev4-1}

\bibliography{RRDPSbib}

\end{document}